\documentstyle[preprint,aps,epsf,floats,psfig]{revtex}

\tighten

\def\OMIT#1{{}}

\def\si{{}^1\kern-.14em S_0}
\def\siii{{}^3\kern-.14em S_1}
\def\piii{{}^3\kern-.14em P_1}
\def\diii{{}^3\kern-.14em D_1}

\begin{document}

\preprint{\vbox{
\hbox{NT@UW-03-06}
\hbox{INT-PUB-03-06}
}}

\title{The $\Lambda_Q \Lambda_Q$ Potential}
\author{{\bf Daniel Arndt}$^a$,
{\bf Silas R.~Beane}$^b$  and {\bf Martin J.~Savage}$^{a}$}
\address{$^a$ Department of Physics, University of Washington, \\
Seattle, WA 98195. }
\address{$^b$ Institute for Nuclear Theory, University of Washington, \\
Seattle, WA 98195. }

\maketitle

\begin{abstract} 

Lattice QCD simulations of the potential between two baryons, each
containing a heavy quark and two light quarks, such as the $\Lambda_Q
\Lambda_Q$ potential, will provide insight into the nucleon-nucleon
interaction.  As one-pion exchange does not contribute to the
$\Lambda_Q \Lambda_Q$ potential, the long-distance behavior is
dominated by physics that contributes to the intermediate-range
attraction between two nucleons. We compute the leading long-distance
contributions to the $\Lambda_Q \Lambda_Q$ potential in QCD and in
partially-quenched QCD in the low-energy effective field theory.

\end{abstract}

\bigskip
\vskip 8.0cm
\leftline{April 2003}

\vfill\eject

\section{Introduction}

The nucleon-nucleon (NN) potential has been studied for many decades,
and modern NN potentials have been constructed to reproduce all NN
scattering data below $\sim 500~{\rm MeV}$ to very-high
precision~\cite{ModPot}.  It is well established that the
long-distance component of the potential is due to one-pion exchange
(OPE). A clear demonstration of this point is provided in work by the
Nijmegen group where the charged and neutral pion masses are free
parameters in the OPE component that are fit to the NN phase-shift
data~\cite{NijOPE}.  The fit values of the masses reproduce with high
accuracy the physical values of the pion masses.  Somewhat more
controversial are the intermediate-range and short-range components of
the NN potential.  In the simplest potentials based on one-meson
exchange (OME) the intermediate-range attraction is supplied by the
exchange of the (fictitious) ``$\sigma$''-meson, while the short-range
repulsion is supplied by the exchange of the $\rho$ and $\omega$
mesons.  In modern potentials, such as the AV$_{18}$ potential, the
intermediate- and short-range components of the NN potential are
not assumed to arise from the exchange of a single meson and {\it
ad hoc } smooth functions are used to best reproduce the NN phase-shift
data~\cite{ModPot}.  The last decade has seen considerable progress
towards formulating NN interactions and multi-nucleon interactions in
an effective field theory (EFT) consistent with the
approximate chiral symmetry of QCD~\cite{EFTrevs}.  In fact, recently
it was shown that an NN potential computed with the EFT better
reproduces the NN scattering data than the potentials based on
OME~\cite{TPEfit}.  More precisely, the two-pion exchange (TPE)
component of the potential provides a better description of the
intermediate-range component than the exchange of a
``$\sigma$''-meson~\cite{TPEfit}.  Further it was shown that the
extracted values of coefficients of one-nucleon local operators in the
chiral Lagrangian that gives rise to the potential are consistent with
their determination in the $\pi N$ sector~\cite{TPEfit}.  This
strongly suggests that the intermediate-distance components of the NN
potential can be described by the exchange of uncorrelated pions with
interactions constrained by chiral symmetry.
This is a very encouraging situation, as ultimately one wants to make
a direct and quantitative connection between QCD and nuclear physics
using the language of effective field theory~\cite{EFTrevs}.

It is through lattice QCD simulations that definitive predictions
in nuclear physics will be made directly from QCD. Given the current
state of technology, simulations of multi-nucleon systems are intractable,
but realistic simulations of two-nucleon systems are presently feasible.
Arguably, the most promising method is to calculate scattering phase shifts
directly using L\"uscher's finite-volume algorithm, which, for instance, expresses the
ground-state energy of a two-particle state as a perturbative expansion in the scattering
length divided by the size of the box~\cite{Luscher}. 
This method has been used successfully to study $\pi\pi$ scattering~\cite{Gupta:1993rn}.
There has also been an attempt to compute NN scattering parameters
in lattice QCD; Ref.~\cite{fuku} computes the S-wave scattering
lengths in quenched QCD (QQCD) using L\"uscher's method. 

A second approach to the NN system on the lattice is to study the
simplified problem of two interacting heavy-light
particles~\cite{Richards:1990xf,Mihaly:1996ue,Stewart:1998hk,Michael:1999nq,Richards:2000ix,Fiebig:2002kg}.
It has been suggested that lattice QCD simulations of the potential
between hadrons containing a heavy quark will provide insight into the
nature of the intermediate-range force between two
nucleons~\cite{JLabMIT}. In the heavy-quark limit, the kinetic energy
of the heavy hadrons is absent and the lowest-lying energy
eigenvalues, which can be measured on the lattice, are given by the
interaction potential. To our knowledge, all discussions to date have
addressed two heavy mesons, say $B B$. This case is somewhat 
complicated by the fact that $B$ and $B^*$ are degenerate in the heavy-quark limit
and therefore require a coupled-channel analysis.
In this work we consider the $\Lambda_Q\Lambda_Q$ interaction (where the
$\Lambda_Q$ is the lowest-lying baryon containing one heavy quark, $Q$), 
which does not suffer from this complication, and we compute the leading contribution to the
$\Lambda_Q\Lambda_Q$ potential in the chiral expansion.
Since the $\Lambda_Q$ is an isosinglet, there is no OPE, and the leading
large-distance behavior is governed by TPE, which is physics analogous to
the intermediate-range attraction in the NN potential.

Presently, unquenched lattice simulations with the physical values of
the light-quark masses are prohibitively time-consuming, even on the
fastest machines.  While some quenched calculations can be performed
with the physical quark masses there is no limit in which they
reproduce QCD, and consequently they should be considered to be a
warm-up exercise for the ``real thing''.  Relatively recently it was
realized that partially-quenched (PQ) simulations, in which the sea
quarks are more massive than the valence quarks, provide a rigorous
way to determine QCD observables and are less time-consuming than
their QCD counterparts.  It is with PQ simulations that
nuclear-physics observables will first be calculated rigorously from
QCD.  As PQQCD reproduces QCD only in the limit in which the sea-quark
masses become equal to the valence-quark masses which, in turn, are
set equal to the physical values of the light quark masses (we call
this the QCD limit), there are some interesting features of the PQ
theory that are distinct from nature away from the QCD limit.  In QCD,
the long-distance component of the NN potential is due to OPE, as
discussed above.  However, in PQQCD there is also a contribution from
the exchange of the $\eta$-meson (in the theory with two flavors of
light quarks).  In QCD such an exchange is suppressed due to the large
mass of the $\eta$ compared to the $\pi$. However, in PQQCD the $\eta$
propagator has a double-pole component that depends on the pion mass
due to the hairpin interactions with a coefficient that depends upon
the difference between the masses of the sea and valence quarks.
Therefore, away from the QCD limit the long-distance component of the
NN potential is dominated by one-eta exchange (OEE) and falls
exponentially with a range determined by the pion mass, $\sim
e^{-m_\pi r}$, as opposed to the familiar Yukawa type
behavior~\cite{BSpot}.  This exponential behavior is even more
pronounced in quenched calculations.  Therefore, it is likely that
quenched and PQ calculations of the NN potential will unfortunately be
dominated by quenching and partial-quenching artifacts.  While this
seems like a negative state-of-affairs, there is hope that the PQ
simulations will tell us something about the NN interaction. An EFT
with consistent power-counting describing the multi-nucleon
sector~\cite{BBSvK} has recently been developed, and its
partially-quenched counterpart was straightforwardly
constructed~\cite{BSPQNN}.  The NN scattering amplitudes in the
S-waves were computed out to next-to-leading order (NLO) in the EFT
power-counting, while the contributions to the higher partial waves,
resulting from OPE and OEE, were computed at leading order.  The hope
is that finite-volume lattice calculations~\cite{Luscher} can be
performed to extract the NN phase-shifts directly from PQQCD through
the use of the scattering amplitudes computed in Ref.~\cite{BSPQNN}.
Unfortunately, given the violence that quenching and partial-quenching
do to the long-distance component of the NN potential, there is little
hope that such calculations can shed light on the nature of the
intermediate-range component of the NN force.

\section{The $\Lambda_Q\Lambda_Q$ Potential in QCD}

The lowest-lying baryons containing a single heavy quark can be
classified by the spin of their light degrees of freedom (dof), $s_l$,
in the heavy-quark limit, $m_Q\rightarrow\infty$.  Working with two
light flavors, $u$ and $d$ quarks, the light dof of the isosinglet
baryon, $\Lambda_Q$, have $s_l=0$, while the light dof of the
isotriplet baryons, $\Sigma_Q^{\pm 1,0}$ and $\Sigma_Q^{\pm 1,0 *}$,
(the superscript denotes the third component of isospin) have
$s_l=1$~\footnote{For three light flavors the baryons fall into a
${\bf 6}\oplus \overline{\bf 3}$ of SU(3).}.  In the heavy-quark limit
the spin-${1\over 2}$ $\Sigma_Q^{\pm 1,0}$ baryons are degenerate with
the spin-${3\over 2}$ $\Sigma_Q^{\pm 1,0 *}$ baryons, but are split in
mass from the $\Lambda_Q$ by an amount that does not vanish in the
chiral limit.  The Lagrange density describing the interactions of
these heavy baryons with the pions is~\cite{ChoHB,YanHB}
\begin{eqnarray}
{\cal L} & = & 
i \overline{\Lambda}_Q v\cdot D \Lambda_Q\ -\ i\overline{S}_\mu^{ij} v\cdot D
S^\mu_{ij}
\ +\  \overline\Delta  \, \overline{S}_\mu^{ij} S^\mu_{ij}
\ +\ 
i g_2 \epsilon_{\mu\nu\sigma\lambda}\ \overline{S}^{\mu,ij} v^\nu
\left(A^\sigma\right)_i^k S^\lambda_{jk}
\nonumber\\
&& \ +\ g_3 \left( \overline{\Lambda}_Q \epsilon^{ij}  
\left(A_\mu\right)_i^k S^{\mu}_{jk}
\ +\ {\rm h.c}
 \right)
\ \ \ ,
\label{eq:hbcl}
\end{eqnarray}
where $S^\mu_{ij}$ is the superfield containing both the
$\Sigma_Q^{\pm 1,0}$ and $\Sigma_Q^{\pm 1,0 *}$, $A_\mu = {i\over
2}\left(\xi\partial_\mu\xi^\dagger -
\xi^\dagger\partial_\mu\xi\right)$ is the axial-vector field of pions,
$v_\mu$ is the heavy-hadron four-velocity, $g_{2,3}$ are coupling
constants that must be determined either from data or from lattice
QCD, and $\overline\Delta$ is the $\Sigma_Q^{(*)}-\Lambda_Q$ mass splitting
in the heavy-quark limit. 
Experimentally, the measured width of the $\Sigma_c^{*++}$ is
$\Gamma (\Sigma_c^{*++}) = 17.9^{+3.8}_{-3.2}\pm 4.0~{\rm
MeV}$~\cite{pdg}, which fixes the axial coupling $g_3$ to be
$|g_3|=0.95\pm 0.08\pm 0.08$ in the charmed sector.  Recently, it has been conjectured that in QCD,
$g_3=1$ and $g_2=2$ in the chiral limit~\cite{BSconj}.  The coupling 
$g_2$ might be measured through the radiative decays of the $\Sigma_c^{(*)}$~\cite{SavageLu}.

\begin{figure}[htb]
\centerline{{\epsfxsize=4.0in \epsfbox{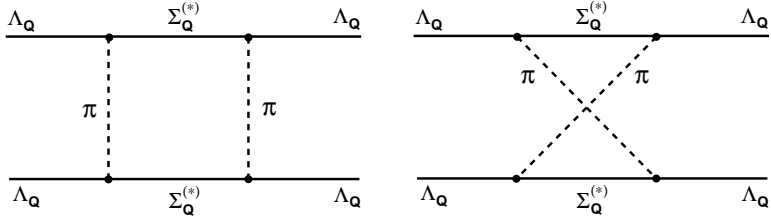}}} 
\vskip 0.15in
\noindent
\caption{\it 
The box and crossed-box diagrams that give the longest-distance component of
the $\Lambda_Q\Lambda_Q$ potential. The vertices are taken from the Lagrangian
in eq.~(\ref{eq:hbcl}).
Heavy-quark symmetry forbids $\Lambda_Q\Lambda_Q$ intermediate states in the
box and crossed-box diagrams.
}
\label{fig:BandCB}
\vskip .0in
\end{figure}

As the light dof in the $\Lambda_Q$ have $s_l=0$ in the heavy-quark
limit, the light-quark axial current matrix element vanishes, and thus
there is no $\Lambda_Q\Lambda_Q\pi$ interaction at leading order in
the heavy quark expansion.  This means that there is no OPE (or OEE)
contribution to the $\Lambda_Q\Lambda_Q$ potential in QCD and PQQCD,
and therefore there is no long-distance component in the
$\Lambda_Q\Lambda_Q$ potential.  It is the two-pion exchange box and
crossed-box diagrams, as shown in Fig.~\ref{fig:BandCB}, that provide
the longest-distance interaction between two $\Lambda_Q$'s. 
In addition, there are local four-$\Lambda_Q$ operators at the same order in the
chiral expansion
but such local interactions give
coordinate-space delta-functions.
Not only are such local interactions allowed, they are required to absorb the
divergences arising in the box and crossed-box diagrams in Fig.~\ref{fig:BandCB}.
Analogous diagrams to those in Fig.~\ref{fig:BandCB}
in the two-nucleon sector provide part of the
intermediate-range component of the NN potential.  
However, it is important to realize
that there are additional interactions that contribute to the
intermediate-range component of the NN potential in the chiral expansion, for
instance contributions from the Weinberg-Tomazawa term and also from 
higher-dimension ${\overline N}N\pi\pi$ vertices.
Therefore, while  the $\Lambda_Q\Lambda_Q$ potential provides a window
into the nature of the intermediate-range NN interaction, it certainly does
not provide a complete description.

\begin{figure}[ht]
\centerline{\psrotatefirst
\psfig{file=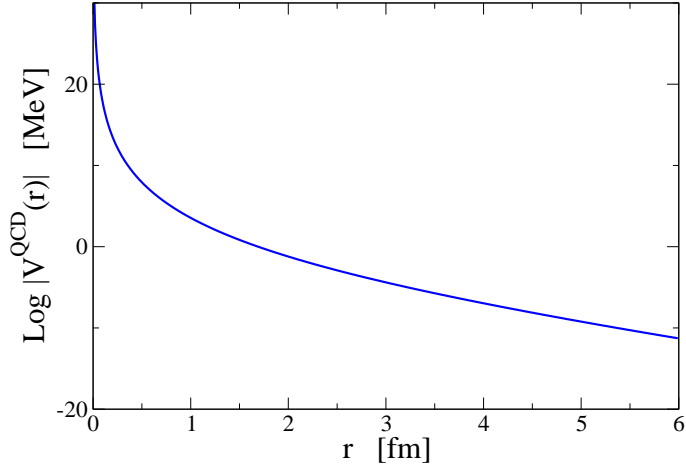,width=4in,angle=-90}
}
\vskip 0.15in
\noindent
\caption{\it 
The logarithm of the $\Lambda_Q\Lambda_Q$ potential in {\rm MeV}
as function of the distance between the $\Lambda_Q$ in {\rm fm}.
}
\label{fig:qcdpot}
\vskip .0in
\end{figure}

If the $\Lambda_Q$ and $\Sigma_Q^{(*)}$ were degenerate,
we would be required to solve the coupled-channel system with 
$\Lambda_Q\Lambda_Q$ and $\Sigma_Q^{(*)}\Sigma_Q^{(*)}$ coupled to $I=0$.
In the charmed sector the $\Sigma_c -\Lambda_c $ mass splitting
is $\Delta=167.1~{\rm MeV}$ and the  $ \Sigma_c^* -\Lambda_c$ mass splitting
is $\Delta=232.7~{\rm MeV}$, and we use the spin-weighted average of these
splittings to estimate
$\overline\Delta\sim 211~{\rm MeV}$ in the heavy-quark limit.
There is no symmetry reason for this mass-splitting to vanish in
the chiral limit, and hence
there is no infrared divergence that
requires a coupled-channel analysis.
In the power-counting we treat 
$\overline\Delta\sim m_\pi$ and take the 
$M_{\Lambda_Q}, M_{\Sigma_Q}\rightarrow \infty$ limit in evaluating the
diagrams in Fig.~\ref{fig:BandCB}.
With this power-counting, one can directly use the Feynman rules of
Heavy-Baryon Chiral Perturbation Theory (HB$\chi$PT) to describe the
low-momentum dynamics of the nucleon and $\Delta$-resonance,
developed by Jenkins and Manohar~\cite{JMHB}, without the need to resum the 
baryon kinetic energy term as is the case for the box and crossed-box 
diagrams in the nucleon sector.
Evaluating the diagrams in Fig.~\ref{fig:BandCB} and then Fourier transforming
them to position space is straightforward.
Using the integral representations of Ref.~\cite{RStricks}, 
the potential between two $\Lambda_Q$'s in QCD
at leading order in the chiral expansion and in the isospin limit is
\begin{eqnarray}
V^{\rm QCD}(r) & = & 
- {g_3^4\over 8\pi^3 f_\pi^4r^2}
\int_0^\infty\ d\lambda\ 
{\overline\Delta^2\over (\lambda^2+\overline\Delta^2)^2}\ 
{\cal I} (\lambda,m_\pi,r)
\ \ \ ,
\end{eqnarray}
where for notational simplicity we have defined the function
${\cal I} (\lambda,m,r)$ to be
\begin{eqnarray}
{\cal I} (\lambda,m,r) & = & 
e^{-2 r \sqrt{\lambda^2+m^2}}
\left[\ 2\left(\lambda^2+m^2 + { 3\sqrt{\lambda^2+m^2}\over r}
+{3\over r^2} \right)^2
\ +\ 
\left(\lambda^2+m^2\right)^2
\ \right]
\ \ \ ,
\end{eqnarray}
and $f_\pi\sim 132~{\rm MeV}$.
A useful feature of this integral representation is the separation 
between the $\overline\Delta$ and $m_\pi$ dependence.
The potential is shown in  Fig.~\ref{fig:qcdpot}.
For asymptotically large distances, the potential is well-approximated by
\begin{eqnarray}
V^{\rm QCD}(r) & \rightarrow & 
- {3 \ g_3^4\  m_\pi^{9/ 2}\over 16\ \pi^{5/ 2}\  r^{5/ 2}\ \overline\Delta^2 f_\pi^4}
e^{-2 m_\pi r}
\ +\ ...
\ \ \ ,
\end{eqnarray}
which exhibits the expected fall off with a length scale set by
twice the mass of the pion and where the dots represent subleading 
contributions in the large-distance expansion.

\section{The $\Lambda_Q\Lambda_Q$ Potential in PQQCD}

The extension of the heavy-baryon sector from QCD to PQQCD is straightforward\footnote{
The extension of the heavy-baryon sector
to quenched QCD was carried out in Ref.~\cite{ChildQ}}.
We use the notation and methods of Ref.~\cite{PQbaryons}.
To establish the representations of $SU(4|2)$ containing the $\Lambda_Q$ and 
$\Sigma_Q^{(*)}$, we consider their interpolating fields.
The interpolating field that contains the $\Lambda_Q$ is
\begin{eqnarray}
{\cal T}_{ij}^\gamma & \sim & 
{\cal Q}^{\gamma,c}\ 
\left[\ 
Q_i^{\alpha,a}Q_j^{\beta,b}\ +\ 
Q_j^{\alpha,a}Q_i^{\beta,b}
\ \right]\ 
\epsilon_{abc} \left(C\gamma_5\right)_{\alpha\beta}
\ \ \ ,
\end{eqnarray}
while the interpolating field that contains the $\Sigma_Q$'s is
\begin{eqnarray}
{\cal S}_{ij}^{\gamma,\mu} & \sim & 
{\cal Q}^{\gamma,c}\ 
\left[\ 
Q_i^{\alpha,a}Q_j^{\beta,b}\ -\ 
Q_j^{\alpha,a}Q_i^{\beta,b}
\ \right]\ 
\epsilon_{abc} \left(C\gamma^\mu\right)_{\alpha\beta}
\ \ \ ,
\end{eqnarray}
where $Q_i^{\alpha,a}$ is a light-quark field operator with graded flavor index
$i=1,2,...,6$ where indices $1,2,3,4$ are fermionic and $5,6$ are bosonic.
The other indices are the color index $a$ and Dirac index $\alpha$.
The heavy-quark field operator is ${\cal Q}^{\gamma,c}$ that does not carry a
graded flavor index.
Under interchange of the light-quark flavor indices we
find that 
\begin{eqnarray}
{\cal T}_{ij}^{\gamma} & = & \left(-\right)^{\eta_i\eta_j} {\cal T}_{ji}^{\gamma}
\ \ ,\ \ 
{\cal S}_{ij}^{\gamma,\mu} \ = \ \left(-\right)^{1+\eta_i\eta_j} {\cal S}_{ji}^{\gamma,\mu}
\ \ \ ,
\end{eqnarray}
where $\eta_i=+1$ for fermionic indices and $\eta_i=0$ for bosonic indices.
It is then straightforward to show, following the techniques of Ref.~\cite{PQbaryons},
that the $\Lambda_Q$ is embedded in a {\bf 17} dimensional representation of
$SU(4|2)$, ${\cal T}_{ij}$,
while the $\Sigma_Q^{(*)}$ are each embedded in a {\bf 19}
dimensional representation of $SU(4|2)$, ${\cal S}_{ij}$.
Under $SU(2)_q\otimes SU(2)_j\otimes SU(2)_{\tilde q}$ the 
ground floor of level A of ${\cal T}_{ij}$ transforms as a $(\bf{1},\bf{1},\bf{1})$ and
contains the $\Lambda_Q$, the first floor of level A contains four baryons
transforming as a $(\bf{2},\bf{1},\bf{2})$, while the ground floor of level B 
contains four baryons that transform as $(\bf{2},\bf{2},\bf{1})$.
Similarly, 
the 
ground floor of level A of ${\cal S}_{ij}$ transforms as a $(\bf{3},\bf{1},\bf{1})$ and
contains the $\Sigma_Q^{(*)}$'s, 
the first floor of level A contains four baryons
transforming as a $(\bf{2},\bf{1},\bf{2})$, while the ground floor of level B 
contains four baryons that transform as $(\bf{2},\bf{2},\bf{1})$.
The remaining members of each irreducible representation do not contribute to 
the potential at the order we are working.
The Lagrange density describing the dynamics of the lowest-lying baryons and the
pseudo-Goldstone bosons is
\begin{eqnarray}
{\cal L} & = &
i \left( \overline{\cal T} v\cdot D  {\cal T} \right)
\ -\ 
i \left( \overline{\cal S}^\mu v\cdot D  {\cal S}_\mu \right)
\ + \ \overline\Delta \left( \overline{\cal S}^\mu {\cal S}_\mu \right)
\nonumber\\
&& \ +\ 
i \ g_2\  \epsilon_{\mu\nu\sigma\lambda}
\ \left(\overline{\cal S}^\mu\ v^\nu\ A^\sigma\ {\cal S}^\lambda\ \right)
\ +\ 
\sqrt{2} \ g_3 \ \left(\overline{\cal T}\ A^\mu \ {\cal S}_\mu\ \right)
\ \ \  ,
\label{eq:superL}
\end{eqnarray}
where the covariant derivatives are
\begin{eqnarray}
D_\mu {\cal T}_{ij} & = & 
\partial_\mu {\cal T}_{ij}
\ +\ \left( V_\mu\right)_i^{\ l} {\cal T}_{lj}
\ +\ \left(-\right)^{\eta_i (\eta_j+\eta_m)}
\left( V_\mu\right)_j^{\ m} {\cal T}_{im}
\ \ \ ,
\end{eqnarray}
\begin{figure}[htb]
\centerline{{\epsfxsize=2.in \epsfbox{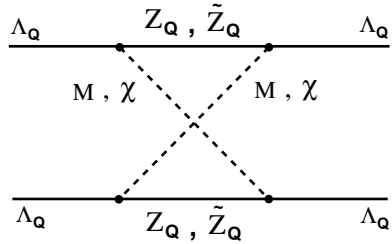}}}
\vskip 0.15in
\noindent
\caption{\it 
Contributions to the crossed-box diagram from mesons and baryons containing
ghost and sea quarks. The vertices are taken from the Lagrangian
in eq.~(\ref{eq:superL}).
The $Z_Q$ are baryons containing one sea quark that transform as a $(\bf{2},\bf{2},\bf{1})$
under $SU(2)_q\otimes SU(2)_j\otimes SU(2)_{\tilde q}$
, while the $\tilde Z_Q$ are baryons containing one ghost quark 
 that transform as a $(\bf{2},\bf{1},\bf{2})$.
The $\chi$ are fermionic mesons
containing one ghost quark while the $M$ are mesons containing one sea quark.
}
\label{fig:PQCB}
\vskip .0in
\end{figure}
and similarly for the ${\cal S}^\mu$.
The index contractions are
\begin{eqnarray}
\left(\overline{\cal T}\ \Gamma\  {\cal T}\right) & = & 
\overline{\cal T}^{\alpha,ji} \ \Gamma_\alpha^\beta \ {\cal T}_{\beta,ij}
\ \ ,\ \
\left(\overline{\cal S}^\mu\ \Gamma\  {\cal S}_\mu\right) \ =\ 
\overline{\cal S}^{\alpha\mu ,ji}\  \Gamma_\alpha^\beta\  {\cal S}_{\beta\mu
  ,ij}
\nonumber\\
\left(\overline{\cal T}\ \Gamma\  Y\ {\cal T}\right) & = & 
\overline{\cal T}^{\alpha,ji} \ \Gamma_\alpha^\beta \ Y_i^l\ 
{\cal T}_{\beta,lj}
\ \ ,\ \
\left(\overline{\cal S}^\mu\ \Gamma\ Y\  {\cal S}_\mu\right) \ =\ 
\overline{\cal S}^{\alpha\mu ,ji}\  \Gamma_\alpha^\beta\ \ Y_i^l\  
{\cal S}_{\beta\mu  ,lj}
\nonumber\\
\left(\overline{\cal T}\ \Gamma\  Y^\mu \ {\cal S}_\mu \right) & = & 
\overline{\cal T}^{\alpha,ji} \ \Gamma_\alpha^\beta \ \left(Y^\mu\right)_i^l\ 
{\cal S}_{\beta\mu,lj}
\ \ \ ,
\end{eqnarray}
where $\Gamma$ acts on the Dirac indices and
$Y\rightarrow UYU^\dagger$ acts in flavor space.

\begin{figure}[ht]
\centerline{\psrotatefirst
\psfig{file=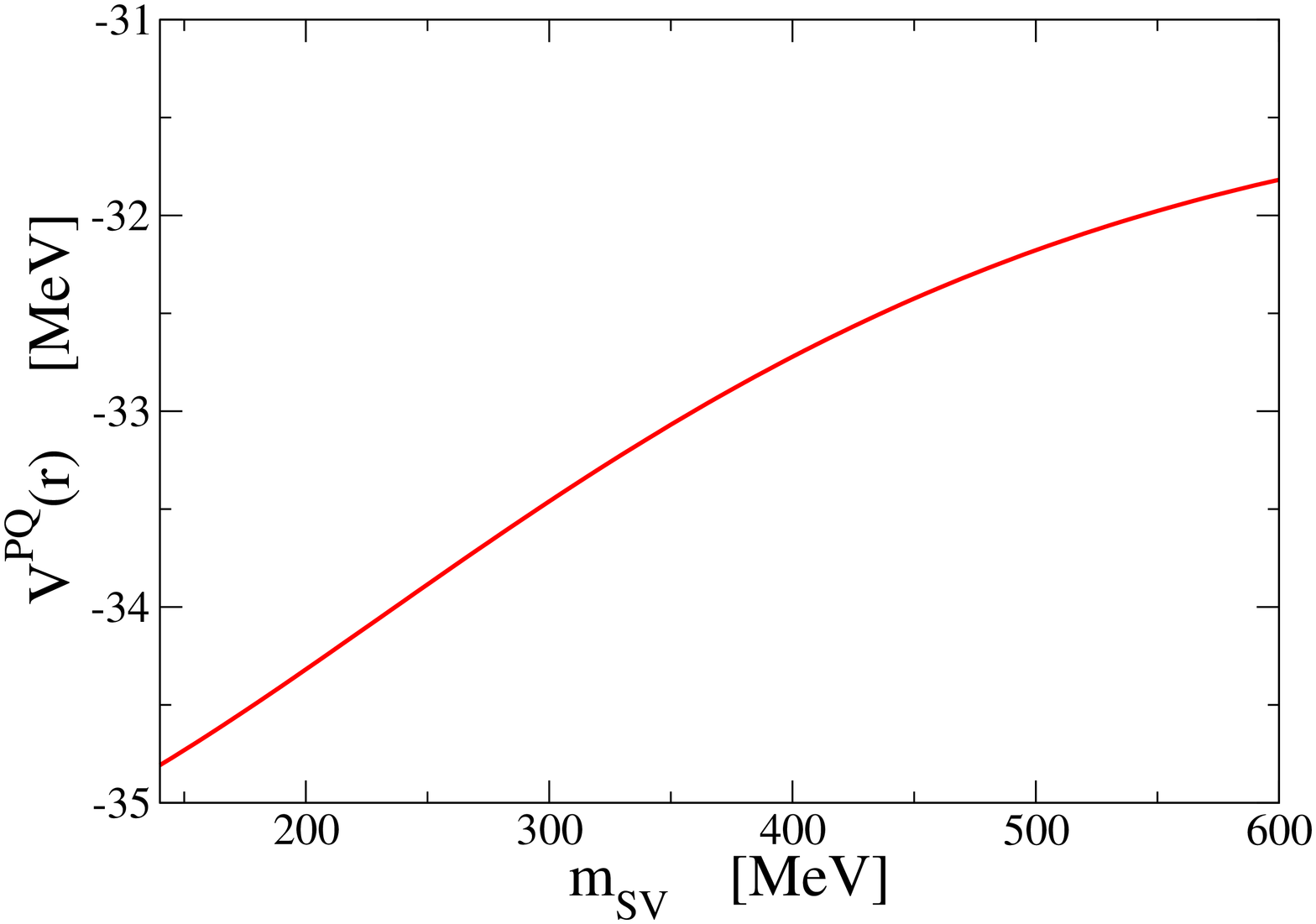,width=3.3in,angle=-90}
\psfig{file=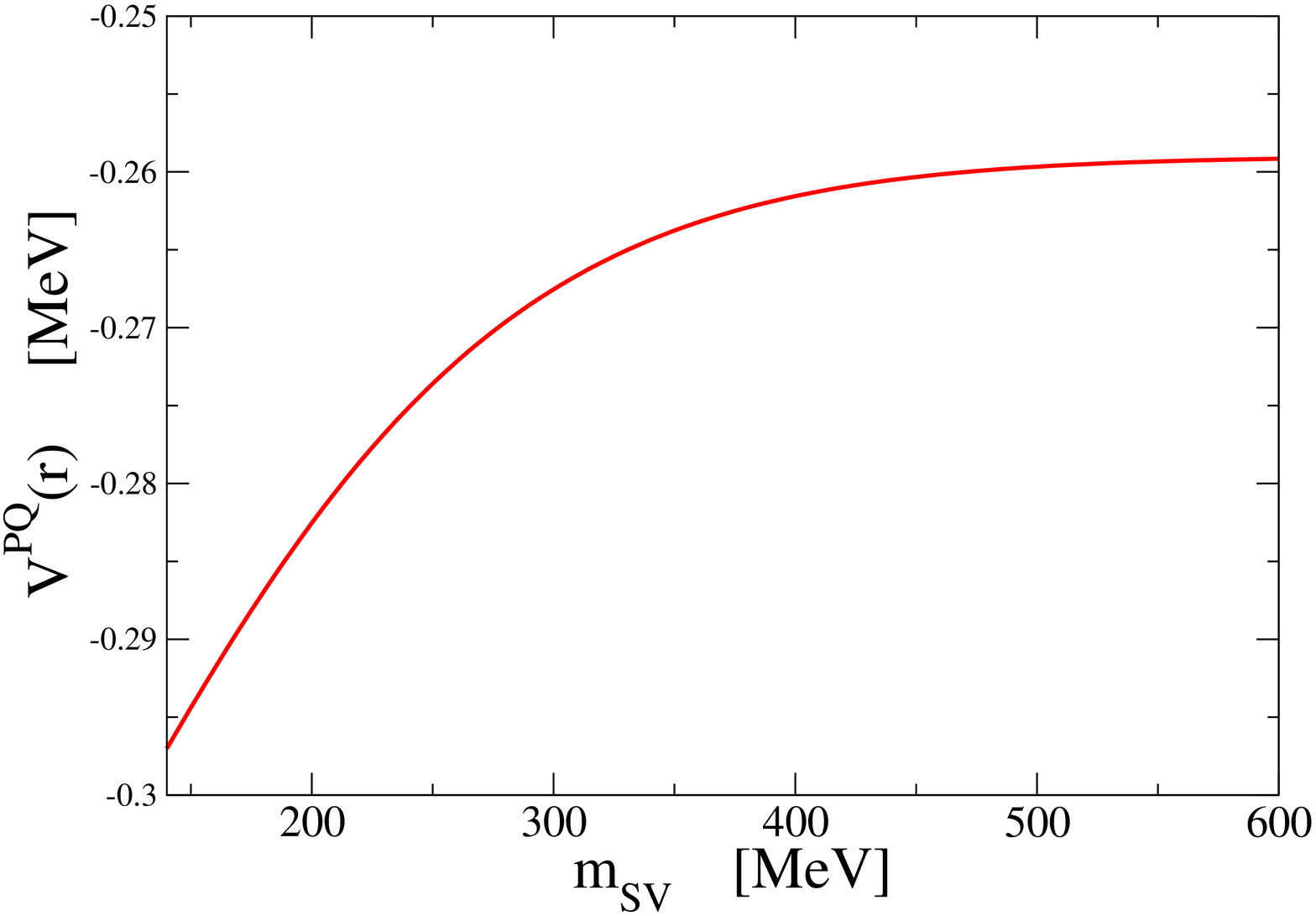,width=3.3in,angle=-90}
}
\vskip 0.15in
\noindent
\caption{\it 
The left panel shows $V^{\rm PQ}(r)$ evaluated at $r=1~{\rm fm}$ 
as a function of the meson mass $m_{SV}$, while the right panels shows 
$V^{\rm PQ}(r)$ evaluated at $r=2~{\rm fm}$.
The vertical axis is in units of ${\rm MeV}$.
When $m_{SV}=m_\pi$ the value of $V^{\rm PQ}(r)$ is equal to 
$V^{\rm QCD}(r)$.
}
\label{fig:Npot12}
\vskip .0in
\end{figure}

In PQQCD the box diagram in Fig.~\ref{fig:BandCB} is unmodified 
due to the fact that one cannot exchange two mesons containing either a ghost
quark or a sea quark and have only two baryons (containing ghost and sea
quarks) in the intermediate state.
However, the crossed-box diagram is modified by quenching and partial
quenching, as shown in Fig.~\ref{fig:PQCB}.
Writing the potential in PQQCD as 
$V^{\rm PQ}(r) = V^{\rm QCD}(r) + \delta V(r)$, we find that
\begin{eqnarray}
\delta V(r) & = & 
+ {g_3^4\over 48\pi^3 f_\pi^4 r^2}\ 
\int_0^\infty\ d\lambda\ {\lambda^2 - \overline\Delta^2\over (\lambda^2+\overline\Delta^2)^2}
\left(\ {\cal I}(\lambda,m_{SV},r)\ -\ {\cal I}(\lambda,m_\pi,r)\ \right)
\ \ \ ,
\end{eqnarray}
where $m_{SV}$ is the mass of a meson containing one sea quark and one
valence quark.
In Fig.~\ref{fig:Npot12} we show $V^{\rm PQ}(r)$ evaluated at 
$r=1~{\rm fm}$ and $r=2~{\rm fm}$ as a function of $m_{SV}$.
In contrast to the NN potential that is dominated at long-distances by 
partial-quenching effects, the $\Lambda_Q\Lambda_Q$ potential 
receives comparable contributions from both the valence quarks and the ghost
quarks, and in fact the (partial) quenching contribution is somewhat smaller than the
QCD contribution.
When $m_{SV}=m_\pi$ we recover the value of the potential in QCD, as has to be
the case.

\begin{figure}[htb]
\centerline{\epsfxsize=2.in \epsfbox{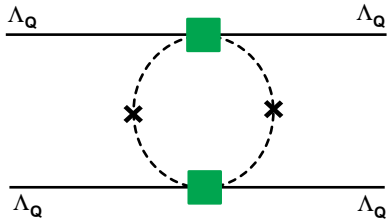}}
\vskip 0.15in
\noindent
\caption{\it The leading hairpin contribution to the $\Lambda_Q\Lambda_Q$ 
potential. The crosses
denote double-pole $\eta$ propagators and the boxes are vertices from 
Lagrange density in eq.~(\ref{eq:hairpin})}
\label{fig:PQCBhp}
\vskip .0in
\end{figure}
Given the large hairpin contribution to the NN potential, it is important to
estimate the size of the hairpin contribution to the $\Lambda_Q\Lambda_Q$ potential.
The Lagrange density for the leading $\eta$ (hairpin) interactions, neglecting
operators involving insertions of the light-quark mass matrix, is
\begin{eqnarray}
{\cal L} & = & {1\over\Lambda_\chi^3}\ 
\left[\ c_1\ \left(\nabla\eta\right)^2\ +\ c_2\ 
  \left(\partial_0\eta\right)^2\ \right]
\ \overline{\Lambda}_Q\Lambda_Q
\ \ \ ,
\label{eq:hairpin}
\end{eqnarray}
where $\Lambda_\chi$ is the chiral-symmetry breaking scale and
$c_{1,2}$ are unknown constants whose natural size is 
$c_{1,2}\sim 1$ by naive dimensional analysis.
One finds that two-insertions of the interactions in eq.~\ref{eq:hairpin},
as shown in Fig.~\ref{fig:PQCBhp}, give a contribution to the 
$\Lambda_Q\Lambda_Q$ potential of
\begin{eqnarray}
V^{\rm HP} (r) & = & 
-{(m_{SS}^2-m_\pi^2)^2\over 32\pi^3 \Lambda_\chi^6}
\int_0^\infty d\lambda\ 
e^{- 2 r \sqrt{\lambda^2+m_\pi^2}}
\ 
\left[ \ 
\left( {2\over r^2} +  \lambda^2+m_\pi^2\right) \ c_1^2 
\ +\ 
2 \lambda^2\  c_1 c_2  
\right.\nonumber\\
& & \left.\qquad\qquad\qquad\qquad\qquad\qquad\qquad\qquad
\ +\ 
c_2^2 \ {\lambda^4\over  {\lambda^2+m_\pi^2}}\ 
\right]
\ \ \ .
\end{eqnarray}
The remaining integral can be performed analytically in terms of modified
Bessel functions and Struve functions, but it is less than illuminating and we
do not present it here.
This contribution is suppressed  compared to the leading non-hairpin contributions as dictated
by the EFT power counting.
Numerically, we find that the hairpin contribution to the potential at
$r=1~{\rm fm}$ is $V^{\rm HP} (1~{\rm fm} ) \sim -1~{\rm keV}$, while at 
$r=2~{\rm fm}$ is $V^{\rm HP} (2~{\rm fm} ) \sim -0.06~{\rm keV}$ for 
$c_1=c_2=1$, $\Lambda_\chi\sim 1~{\rm GeV}$ and $m_{SS}=600~{\rm MeV}$.
At asymptotically large distances the potential becomes
\begin{eqnarray}
V^{\rm HP} (r) & \rightarrow & 
- c_1^2 \ 
{(m_{SS}^2-m_\pi^2)^2  \ m_\pi^{5/2}\over 64\ \pi^{5/ 2}\  
\Lambda_\chi^6} \ 
{e^{-2m_\pi r}\over\sqrt{r}}
\ +\ \ldots
\ \ \ .
\end{eqnarray}
where the dots represent subleading contributions in the large-distance
expansion.
While at asymptotically large distances this contribution is larger than that
from the box and crossed-box diagrams, asymptopia finally sets in at distances at which
all contributions are numerically insignificant.

\section{Conclusions}

In this work we have computed the potential between two $\Lambda_Q$'s at
leading order in effective field theory in both QCD and PQQCD. 
Further, we have
estimated the size of the leading contribution from hairpin interactions in PQQCD.
We find that the partially-quenched $\Lambda_Q\Lambda_Q$ potential
does not suffer from some of the (partial-) quenching problems that plague the
NN potential due to the absence of single pseudo-Goldstone exchange.
The potentials we have computed in this work will allow for the chiral
extrapolation of lattice calculations performed with unphysically large sea
quark masses. 

As these potentials fall off with a mass scale set by $\sim 2 m_\pi$, they
are quite small for baryon separations greater than $r\sim 1.5~{\rm fm}$.
Therefore, the theoretical advantages of studying this system to learn about
the NN potential may be undermined by the difficulties in extracting a signal
from lattice simulations.
However, the simplifications introduced by only having two light quarks, and
a single infinitely-massive quark to fix the inter-baryon separation makes
this system a prime candidate for studying inter-baryon interactions.
We look forward to the Jlab/MIT group~\cite{JLabMIT} carrying 
out their program in this area.

\acknowledgements

We are grateful to Steve Sharpe for providing valuable insight.
This work is supported in part by the U.S. Dept. of Energy under Grant No.~DE-FG03-97ER4014
(D.A. and M.J.S.) and Grant No.~DE-FG03-00-ER-41132 (S.R.B.).


\begin{references}

\bibitem{ModPot} V.G.~Stoks, R.A.~Klomp, C.P.~Terheggen and J.J.~de Swart,
{\it Phys. Rev.} {\bf C49}, 2950 (1994);
R.B.~Wiringa, V.G.~Stoks and R.~Schiavilla,
{\it Phys. Rev.} {\bf C51}, 38 (1995);
R.~Machleidt,
{\it Phys. Rev.} {\bf C63}, 024001 (2001).

\bibitem{NijOPE} M.M.~Nagels, T.A.~Rijken and J.J.~de Swart,
{\it Phys. Rev.} {\bf D12}, 744 (1975).

\bibitem{EFTrevs} 
P.F.~Bedaque and U.~van Kolck,
{\it Ann. Rev. Nucl. Part. Sci.} {\bf 52}, 339 (2002);
D.R.~Phillips, {\it Czech. J. Phys.} {\bf  B49}, 52 (2002);
S.R.~Beane, P.F.~Bedaque, W.C.~Haxton, 
D.R.~Phillips and  M.J.~Savage,
Essay for the Festschrift in honor of Boris Ioffe, 
in the Encyclopedia of Analytic QCD, 
At the Frontier of Particle Physics, vol.~1, 133-271, edited by M.~Shifman
(World Scientific).

\bibitem{TPEfit} 
M.C.~Rentmeester, R.G.~Timmermans, J.L.~Friar and J.J.~de Swart,
{\it Phys. Rev. Lett.}  {\bf 82}, 4992 (1999);
M.C.~Rentmeester, R.G.~Timmermans and J.J.~de Swart,
{\tt nucl-th/0302080}.

\bibitem{Luscher}
M.~L\"uscher,
{\it Commun. Math. Phys.} {\bf 105}, 153 (1986).

\bibitem{Gupta:1993rn}
R.~Gupta, A.~Patel and S.~R.~Sharpe,
{\it Phys. Rev.} {\bf D48}, 388 (1993).

\bibitem{fuku}
M.~Fukugita, Y.~Kuramashi, M.~Okawa, H.~Mino and A.~Ukawa,
{\it Phys. Rev.} {\bf D52}, 3003 (1995).

\bibitem{Richards:1990xf}
D.G.~Richards, D.K.~Sinclair and D.W.~Sivers,
{\it Phys. Rev.} {\bf D42}, 3191 (1990).

\bibitem{Mihaly:1996ue}
A.~Mihaly, H.R.~Fiebig, H.~Markum and K.~Rabitsch,
{\it Phys. Rev.} {\bf D55}, 3077 (1997).

\bibitem{Stewart:1998hk}
C.~Stewart and R.~Koniuk,
{\it Phys. Rev.}  {\bf D57}, 5581 (1998).

\bibitem{Michael:1999nq}
C.~Michael and P.~Pennanen  [UKQCD Collaboration],
{\it Phys. Rev.} {\bf D60}, 054012 (1999).

\bibitem{Richards:2000ix}
D.G.~Richards,
{\tt nucl-th/0011012}.

\bibitem{Fiebig:2002kg}
H.R.~Fiebig and H.~Markum,
{\tt hep-lat/0212037}.

\bibitem{JLabMIT} http://www.jlab.org/\~{\,}dgr/lhpc/march00.pdf;\hfill
http://www.jlab.org/\~{\,}dgr/lhpc/sdac\_proposal\_final.pdf.

\bibitem{BSpot} S.R.~Beane and M.J.~Savage,
{\it Phys. Lett.} {\bf B535}, 177 (2002).

\bibitem{BBSvK} S.R.~Beane, P.F.~Bedaque, M.J.~Savage and U.~van Kolck,
{\it Nucl. Phys.} {\bf A700}, 377 (2002).

\bibitem{BSPQNN} S.R.~Beane and M.J.~Savage,
{\it Phys. Rev.} {\bf D67}, 054502 (2003).

\bibitem{ChoHB} P.L.~Cho,
{\it Phys.  Lett.} {\bf  B285}, 145 (1992);
{\it Nucl. Phys.} {\bf B396}, 183 (1993);
Erratum-ibid. {\bf B421}, 683 (1994).

\bibitem{YanHB} T.-M.~Yan, H.-Y.~Cheng, C.-Y.~Cheung, G.-L.~Lin, Y.C.~Yin and H.-L.~Yu,
{\it Phys. Rev.} {\bf D46} 1148 (1992).

\bibitem{pdg} K.~Hagiwara {\it et al.}  [Particle Data Group Collaboration],
{\it Phys. Rev.} {\bf D66}, 010001 (2002).

\bibitem{BSconj} S.R.~Beane and M.J.~Savage,
{\it Phys. Lett.} {\bf B556}, 142 (2003).

\bibitem{SavageLu} M.~Lu, M.J.~Savage and J.~Walden,
{\it Phys. Lett.} {\bf B369}, 337 (1996).

\bibitem{JMHB} E.~Jenkins and A.V.~Manohar,
{\it Phys. Lett.} {\bf B255}, 558 (1991).

\bibitem{RStricks} T.A.~Rijken and V.G.~Stoks,
{\it Phys. Rev.} {\bf C46}, 73 (1992).

\bibitem{ChildQ}  G.~Chiladze,
{\it Phys. Rev.} {\bf D57}, 5586 (1998).

\bibitem{PQbaryons} J.N.~Labrenz and S.R.~Sharpe,
{\it Phys. Rev.}  {\bf D54}, 4595 (1996);
M.J.~Savage.
{\it Nucl. Phys.} {\bf A700} 359 (2002);
J.-W.~Chen and M.J.~Savage,
{\it Phys. Rev.} {\bf D65}, 094001 (2002); 
S.R.~Beane and M.J.~Savage,
{\it Nucl. Phys.} {\bf A709}, 319 (2002).

\end{references}
\end{document}